\documentclass{IEEEtran4PSCC}

\usepackage{caption}
\usepackage{subcaption}

\usepackage[cmex10]{amsmath}
\usepackage[utf8]{inputenc}
\usepackage{soul}
\usepackage{multirow}
\usepackage{soul,color}

\usepackage{algorithm}
\usepackage[noend]{algpseudocode}

\usepackage{graphicx}
\usepackage{dsfont}
\usepackage{cite}
\usepackage[hidelinks]{hyperref}

\usepackage[utf8]{inputenc}
\usepackage{mathtools}
\usepackage[mathscr]{euscript}
\usepackage{amsmath}
\usepackage{amssymb}
\usepackage{amsfonts}
\usepackage{amssymb}
\usepackage{amsmath}
\usepackage{verbatim}
\usepackage[T1]{fontenc}
\usepackage[utf8]{inputenc}
\usepackage{latexsym}
\usepackage{enumerate}
\usepackage{indentfirst}
\usepackage{calligra}
\usepackage{ulem}
\usepackage{graphicx}
\usepackage{array}
\usepackage{float}
\usepackage{bbm}
\usepackage{tikz}

\begin{document}
%
\title{Flexible and curtailable resource activation in three-phase unbalanced distribution networks}

 \author{\IEEEauthorblockN{Md Umar Hashmi,
 Arpan Koirala,
 Hakan Ergun, 
 Dirk Van Hertem}
 \IEEEauthorblockA{\textit{Electa-ESAT, KU Leuven \& EnergyVille},
Genk, Belgium}
 \IEEEauthorblockA{(mdumar.hashmi, arpan.koirala, hakan.ergun, dirk.vanhertem)@kuleuven.be}
 }
 \makeatletter
\let\old@ps@headings\ps@headings
\let\old@ps@IEEEtitlepagestyle\ps@IEEEtitlepagestyle
\def\psccfooter#1{%
    \def\ps@headings{%
        \old@ps@headings%
        \def\@oddfoot{\strut\hfill#1\hfill\strut}%
        \def\@evenfoot{\strut\hfill#1\hfill\strut}%
    }%
    \def\ps@IEEEtitlepagestyle{%
        \old@ps@IEEEtitlepagestyle%
        \def\@oddfoot{\strut\hfill#1\hfill\strut}%
        \def\@evenfoot{\strut\hfill#1\hfill\strut}%
    }%
    \ps@headings%
}

\makeatother

\psccfooter{%
        \parbox{\textwidth}{\hrulefill \\ \small{22nd Power Systems Computation Conference} \hfill \begin{minipage}{0.2\textwidth}\centering \vspace*{4pt} \includegraphics[scale=0.06]{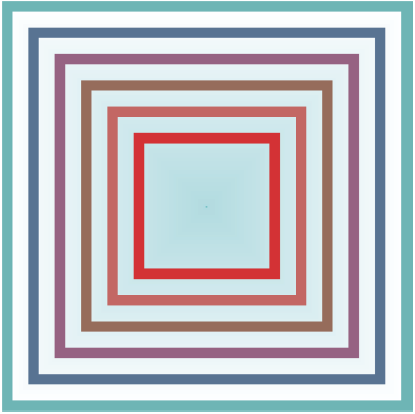}\\\small{PSCC 2022} \end{minipage} \hfill \small{Porto, Portugal --- June 27 -- July 1, 2022}}%
}

\maketitle

\begin{abstract}
The need for flexibility and curtailable resources is crucial for ensuring the healthy operation of future distribution networks (DN).
In this work, we propose a network-state driven framework that distribution system operators (DSOs) can utilize for activating flexible and curtailable resources for alleviating network voltage and thermal issues, while accounting for network voltage and current imbalances.
This approach assumes the availability of dynamic network state information and uses nodal sensitivities for calculating a flexibility activation signal (FAS). The signal design is motivated by volt-Var and volt-watt inverter control, and thus bounded.
The FAS also considers network voltage and current imbalances and incentivizes activation of active and reactive power flexibilities for reducing imbalance in addition to mitigating voltage and thermal imbalances in a three-phase unbalanced distribution network.
The FAS design resembles optimal power flow duals, often used as locational marginal prices.
The gains associated with the imbalance component of the objective function of three-phase unbalanced resource activation (TPU-RA) is performed using Pareto optimality.
A numerical case study is presented showing the efficacy of the proposed framework in avoiding network issues while reducing voltage unbalance factor by more than 80\%.
Further, DN's flexibility needs are quantified for location and time of day.
\end{abstract}
\vspace{-4pt}
\section{Introduction}
With growing distribution network (DN) uncertainty due to new loads and distributed generation (DG), the need for flexible resources is going to be critical for its reliable operation. Prior works on demand response, congestion management, low voltage regulation, peak demand shaving, storage control, provides a huge pool of literature as a starting point for operating flexible resources. 
Fast responding resources which can ramp up and ramp down are needed to damp DN fluctuations \cite{nosair2015flexibility}.
System operators should plan new flexible resources with the growth of DGs and new loads such as electric vehicles \cite{meibetaner2019co}, which can compose more than 50\% of typical residential load.

Flexible resources can increase the reliability of DN supply \cite{sperstad2020impact, klyapovskiy2019incorporating, chen2019aggregate, salpakari2016optimal, kalisch2019assessment}.
Authors in \cite{sperstad2020impact} distinguish security of supply based on energy and power capacity adequacy, reliability of supply and power quality.
Similar to our work, authors in \cite{klyapovskiy2019incorporating} use the value of flexibility to distinguish different resources in a DN. The key difference is our flexibility activation signal (FAS) design is purely network state-driven and does not depend on the source of flexibility.
As such, the focus of this paper ties with the decision taking process at the DSO.
Authors in \cite{salpakari2016optimal} utilize prosumer flexibilities for increasing PV self-consumption, which leads to saving of up to 25\%.
Authors in \cite{huang2017real} utilize flexible demand swaps for balancing in real-time.

We utilize nodal voltage sensitivity in order to bring the locational aspect in the FAS design.
Voltage sensitivities are widely used in many power system applications, such as
battery management \cite{hashemi2014scenario},
inverter operation \cite{zad2018new}.
We use the \textit{perturb-and-observe} method, which approximates sensitivity components for the nodal voltage.
In this work, we use voltage sensitivity towards active and reactive power to design priorities for FAS design in a DN.
Not all flexible and curtailable resources are the same, depending on the location of that resource in the network and network congestion issues at that location. 
The DSO needs a valuation and activation mechanism for resource activation, ensuring new network issues are not created while prevailing network issues are mitigated.

The FAS design is motivated by volt-watt and volt-Var inverter control. Volt-watt and volt-Var
inverter control in standalone and/or in combination are popular \cite{weckx2014combined}, \cite{karagiannopoulos2017hybrid}. It uses drooping behavior to limit excess active (reactive capacitive or inductive) power injection, 
thus not aggravating nodal voltage any further.
This inverter control uses a permissible limit, beyond which it provides proportional active and/or reactive power correction.
Similar to these inverter control policies, our proposed FASs are active in case of DN parameters exceed a permissible level, thus incentivizing resource activation (RA).

DN unbalance is growing with accelerated installations of new consumer loads and DG, which are connected to the network using a single-phase connection. 
{Authors in \cite{chen2019aggregate} utilize distribution network flexibilities for supporting the transmission network while considering an unbalanced DN model. It is crucial to consider an unbalanced DN model as most DN loads are single-phase and thus flexibilities derived from such loads are also single phase. A haphazard resource activation could thus aggravate DN imbalance.
A three-phase analysis is essential to provide adequate signals to single-phase loads in urban grids.}
Authors in \cite{hashmi2020towards} present real-world case studies of an EV charging facility in Pasadena in California and a sub-station on the island of Madeira in Portugal, showing single-phase loads or generation could lead to an increased amount of voltage and current unbalance.
Such an imbalance leads to inefficient use of DN, induction motor overheating and de-rating, increased DN losses on the phase and in the neutral and ground, excess transformer losses, nuisance tripping of relays, reduction in life of consumer appliances to name a few \cite{ma2020review}.
There are passive and active ways of imbalance mitigation \cite{weckx2013phase}, in this work we use flexible resources for reducing voltage and current imbalances in DN.
Unbalance over-compensation is solved by including imbalance components in the TPU-RA objective function.
This work extends our prior work, \cite{hashmi2021sest}, on balanced DN.

\subsection{Contributions, observations, and model description} \vspace{-2pt}
We propose a framework to value flexible resources based on forecast and/or instantaneous network state, i.e, nodal voltage magnitudes, and line loadings. This proposed FAS is utilized to solve DN issues while minimizing the network state-driven resource activation cost. The resource activation cost for a node at a particular time is proportional to the product of nodal FAS and corresponding nodal flexibility activated for a given time.
The FAS is composed of three components, depending on the network state:\\
$\bullet$ \textit{Voltage component}: this is active only if the (measured or forecasted) voltage magnitude is outside the permissible level, \\
$\bullet$ \textit{Thermal loading projection component}: the thermal loading of a branch is projected on network nodes to obtain a nodal activation signal. The nodal projection considers the amount and power flow direction of all branches connected to a node.\\
$\bullet$ \textit{Imbalance component}: In a practical three-phase DN with many single-phase connections, flexibility should be activated,  considering its impact on the distribution network imbalance \cite{douglass2016voltage}.  
    The imbalance component is derived using nodal voltage and projected current imbalance.\\
The voltage and thermal components are motivated by volt-Var and volt-watt inverter control. The imbalance component captures the voltage and current imbalances, which are expected to grow due to the haphazard growth of new single-phase loads and distributed generation.
We observe that the proposed FAS design has some similarities with optimal power flow (OPF) duals, often used as locational marginal prices (LMP) \cite{conejo2005locational}. The proposed FAS provides more information than the OPF duals due to the droop based design. On the contrary, the OPF duals are only active (non-zero) in case of some network constraint violations. 
Using numerical simulations, we show that the proposed resource activation framework for an unbalanced DN not only leads to mitigation of network issues, but also reduced voltage and current imbalance. From grid rules, we observe that minor imbalances in voltage and/or current are not a major concern for a DSO, with this aspect in mind we propose a method to tune the imbalance components of the three-phase unbalance resource activation (TPU-RA) using Pareto-optimality, so as imbalances can be reduced while ensuring the cost of reducing DN imbalance is not too high.

The objective of the resource dispatch problem, TPU-RA, 
{is to 
minimize flexible and curtailable resource activation cost.
The flexibility activation cost is equal to the sum of the product of proposed FAS and the amount of activated flexibility for all the nodes and time instances.}
The associated constraints include traditional optimal power flow constraints such as voltage limit, thermal limit,  nodal power balance, Ohm's law, and generation limit constraints. In addition to these, flexibility and curtailment power limits are included.
In order to have the resource dispatch problem tractable, all associated FASs are generated a priori for performing the main optimization discussed earlier.
This decomposition of the optimization 
avoids the use of 
binary variables in TPU-RA.
The resource activation implementation in TPU-RA has a hierarchical structure.
The cost of activation for flexible resources are set higher than the electricity tariff, which  will incentivize prosumers to opt for using their less priority loads as flexible resources.
In case flexible resources are not enough, the proposed TPU-RA activates curtailment of load and/or generation. 

The paper is organized as follows. 
Section~\ref{sectionfas} presents the model used for calculating FAS.
Section~\ref{sectionOPT} formulates the optimization problem for TPU-RA.
Numerical results are presented in section \ref{sectionNUM} and
section~\ref{sectionCON} concludes the paper.

\section{Flexibility activation signal (FAS)}
\label{sectionfas}

\subsection{Notation}
A power network is composed of several components such as nodes,
branches, generators and loads. 
A network is characterized by $<\mathscr{N},E>$, where $\mathscr{N}$ denotes all nodes in all phases and $E$ denotes branches connecting a pair of nodes. 
Each node $i \in \mathscr{N}$ have three phases denoted by $\phi \in \{A,B,C\}$.
Each node $i$, phase $\phi$ and time $t$ has two variables, i.e., voltage magnitude ($V_{{\phi},i,t}$) and phase angle ($\theta_{{\phi},i,t}$) which are governed by power injection and load magnitude. 
{The branch admittance for phase $\phi$ and branch $(i,j) \in E$ is denoted as $Y_{\phi,ij}$, which governs power flow and losses.}
Nodes with loads connected is denoted as $\mathscr{N}_d \subset \mathscr{N}$. These nodes have active and reactive power loads denoted as $P^d_{{\phi},{i,t}}$ and $Q^d_{{\phi},{i,t}}$.
Nodes with generators connected is denoted as $\mathscr{N}_G \subset \mathscr{N}$, have active and reactive power generation denoted as $P^g_{{\phi},{i,t}}$ and $Q^g_{{\phi},{i,t}}$.
$\mathbbm{1}_{(\text{condition})}$ returns 1 if the condition is true.

\subsection{Flexibility activation signal design}
Flexibilities are assumed to be located at the nodes where load and/or distributed generation is connected.
This is a realistic assumption as flexible resources are derived from \cite{chen2018distributed}
(a) low priority, temporally flexible, deadline constrained loads such as dishwasher, pool pumps,
(b) operational dead-band based prosumer loads such as  thermostatically controlled devices \cite{koch2011modeling},
(c) ramp rate, power and energy-constrained sources such as energy storage \cite{buvsic2017distributed},
(d) curtailable generation and load.
Thus, nodal sensitivities and FAS are only calculated for nodes with load and/or generation connected to it.
{The nodal voltage sensitivity (NVS) is utilized to bound the FAS. The details of NVS calculations are detailed in Appendix \ref{appendix:nodalVs}.}

The efficient activation of distributed flexible and curtailable resources are crucial for the reliable operation of DN. 
The flexibility activation signal (FAS) design is based on the network state, i.e., nodal voltage magnitudes and thermal loading of lines.
As the FAS is nodal, we propose a proportional line parameter projection mechanism that takes into consideration the flow direction, assuming the flow {convention} from the substation to loads is positive.
{The line parameter projection on to nodes is detailed in Appendix \ref{sec:nodalloadingprojection}.}
The active power flexible resources are considered as ramp up and ramp down flexible resources.
Ramp up flexible resources increase DN load and is analogous to generation curtailment.
Similarly, ramp down flexible resources decreases DN load and are analogous to load curtailment.
The reactive power flexible resources are considered capacitive and inductive.
The proposed FAS design for active power are shown in Fig. \ref{fig:componentfle}.

The active power FAS consists of four components: (a) voltage, (b) thermal, (c) voltage imbalance, and (d) current imbalance.
In Fig. \ref{fig:fsacomponent1} and Fig. \ref{fig:fsacomponent2}, the design is motivated by volt-watt inverter control \cite{kashani2018smart}.
{The imbalance components for P and Q are shown in Fig. \ref{fig:fsacomponent3} and Fig. \ref{fig:fsacomponent4} are detailed in Section \ref{subimbalance}.}
{For voltage component of FAS}, the active power consumption is 
{incentivized} if the voltage exceeds the permissible level of $1 + \Delta V_{\text{perm}}$. The saturation level of FAS denoted as $-\text{VC}^{\max}_{{\phi},i,P}$.
Similarly, the active power curtailment is 
{preferred} if voltage dips below the permissible level of $1 -\Delta V_{\text{perm}}$. The saturation level of FAS denoted as $\text{VC}^{\max}_{{\phi},i,P}$.
The thermal component of FAS saturates at $\text{TC}^{\max}_{\phi,i,P}$. The values of these saturating levels are a function of nodal voltage sensitivities of active power for phase $\phi$ and node $i$ denoted as $\text{NVS}_{{\phi},i}^P$.
\begin{equation}\begin{split}
    \text{VC}^{\max}_{\phi,i,P} = f_P(\text{NVS}_{{\phi},i}^P),~~~\text{TC}^{\max}_{\phi,i,P} = g_P(\text{NVS}_{{\phi},i}^P)
    \label{eq:sensitivityComponetFlexPriceP_volt}
\end{split}\end{equation}

The reactive power FAS consists of three components: (a) voltage, (b) thermal and (c) voltage imbalance. The design of reactive power FSA is motivated by volt-Var inverter control \cite{olivier2015active, ustun2019impact}. Here, the reactive power capacitive injection is promoted if voltage dips below the permissible level of $1 -\Delta V_{\text{perm}}$. The saturation level of FAS denoted as $\text{VC}^{\max}_{{\phi},i,Q}$.
The thermal component of FAS saturates at $\text{TC}^{\max}_{\phi,i,Q}$. The values of these saturating levels are a function of nodal voltage sensitivities of reactive power for phase $\phi$ and node $i$ denoted as $\text{NVS}_{{\phi},i}^Q$.
\begin{equation}\begin{split}
    \text{VC}^{\max}_{\phi,i,Q} = f_Q(\text{NVS}_{{\phi},i}^Q),~~~\text{TC}^{\max}_{\phi,i,Q} = g_Q(\text{NVS}_{{\phi},i}^Q)
    \label{eq:sensitivityComponetFlexPriceQ_volt}
\end{split}\end{equation}

{Note $f_P, f_Q, g_P, g_Q$ are linear functions of NVS. Thus,
\eqref{eq:sensitivityComponetFlexPriceP_volt} and \eqref{eq:sensitivityComponetFlexPriceQ_volt} denotes the value of a flexible resource located at a node $i$ is proportional to the NVS of P and Q of the node. The voltage and current imbalance components are linearly related to the DN imbalance.}
All FAS components are detailed next.

%


\subsection{Voltage component of FSA}
\begin{figure}
    \centering
    
        \begin{subfigure}[b]{0.24\textwidth} 
	    \begin{tikzpicture}[line width=0.8pt, =\tikzscale]
	\definecolor{gg}{RGB}{204,255,153}
	\definecolor{orang}{RGB}{255,192,182}
	\draw[draw = none, fill = gg] (0.0,0.0) rectangle (2,-1.5);
	\draw[draw = none, fill = orang] (0.0,0.0) rectangle (-2,1.5);
	\draw[->, line width = 1pt] (-2.2, 0) -- (2.2,0);
	\draw[->, line width = 1pt] (0, -1.9) -- (0,1.9);
	\draw[line width = 0.8pt] (-2, 1.4) -- (-1.8,1.4);
	\draw[line width = 0.8pt] (-1.8, 1.4) -- (-1,0);
	\draw[line width = 0.8pt] (-1, 0) -- (1,0);
	\draw[line width = 0.8pt] (1, 0) -- (1.8,-1.4);
	\draw[line width = 0.8pt] (1.8, -1.4) -- (2,-1.4);
	\draw (0.1,1.8) node [anchor=west] {$(\lambda^{flex,P}_{\phi,i,t})_{\text{voltage}}$};
	\draw[line width = 0.6pt, dotted] (-1.8, 1.4) -- (0,1.4);
	\draw[line width = 0.6pt, dotted] (-1.8, 1.4) -- (-1.8,0);
	\draw[line width = 0.6pt, dotted] (1.8, -1.4) -- (0,-1.4);
	\draw[line width = 0.6pt, dotted] (1.8, -1.4) -- (1.8,0);
	\draw (1.8,0.3) node [anchor=center] {$V_{\max}$};
	\draw (-0.8,1.15) node [anchor=center] {\small{$VC^{\max}_{\phi,i,P}$}};
	\draw (0.8,-1.15) node [anchor=center] {\small{$-VC^{\max}_{\phi,i,P}$}};
	\draw (-1.8,-0.3) node [anchor=center] {$V_{\min}$};
	\draw (0.0,0.4) node [anchor=west] {\footnotesize{$\Delta V_{\text{perm}}$}};
	\draw[<->, line width = 0.5pt] (0, 0.1) -- (1,0.1);
	\draw (0,1.3) node [anchor=west] {Voltage FSA};
	\draw (0,0.9) node [anchor=west] {component};
\end{tikzpicture}
        \caption[case_1]{Voltage component of FAS}    
        \label{fig:fsacomponent1}
    \end{subfigure}
        \begin{subfigure}[b]{0.24\textwidth} 
	    \begin{tikzpicture}[line width=0.8pt, =\tikzscale]
	\definecolor{gg}{RGB}{204,255,153}
	\definecolor{orang}{RGB}{255,192,182}
	\draw[draw = none, fill = gg] (0.0,0.0) rectangle (2,1.5);
	\draw[draw = none, fill = orang] (0.0,0.0) rectangle (-2,-1.5);
	\draw[->, line width = 1pt] (-2.2, 0) -- (2.2,0);
	\draw[->, line width = 1pt] (0, -1.9) -- (0,1.9);
	\draw[line width = 0.8pt] (-2, -1.4) -- (-1.8,-1.4);
	\draw[line width = 0.8pt] (-1.8, -1.4) -- (-1,0);
	\draw[line width = 0.8pt] (-1, 0) -- (1,0);
	\draw[line width = 0.8pt] (1, 0) -- (1.8,1.4);
	\draw[line width = 0.8pt] (1.8, 1.4) -- (2,1.4);
	\draw (0,1.8) node [anchor=west] {$(\lambda^{flex,P}_{\phi,i,t})_{\text{thermal}}$};
	\draw[line width = 0.6pt, dotted] (-1.8, -1.4) -- (0,-1.4);
	\draw[line width = 0.6pt, dotted] (-1.8, -1.4) -- (-1.8,0);
	\draw[line width = 0.6pt, dotted] (1.8, 1.4) -- (0,1.4);
	\draw[line width = 0.6pt, dotted] (1.8, 1.4) -- (1.8,0);
	\draw (1.8,-0.3) node [anchor=center] {$100\%$};
	\draw (0.8,1.15) node [anchor=center] {\small{$TC^{\max}_{\phi,i,P}$}};
	\draw (-0.8,-1.15) node [anchor=center] {\small{$-TC^{\max}_{\phi,i,P}$}};
	\draw (-1.8,0.3) node [anchor=center] {$-100\%$};
	\draw (0.0,0.4) node [anchor=west] {\footnotesize{$\Delta T_{\text{perm}}$}};
	\draw[<->, line width = 0.5pt] (0, 0.1) -- (1,0.1);
	\draw (0,1.7) node [anchor=east] {Thermal FSA};
	\draw (0,1.3) node [anchor=east] {component};
\end{tikzpicture}
        \caption[case_2]{Thermal component of FAS}    
        \label{fig:fsacomponent2}
    \end{subfigure}
    
       \vskip \baselineskip
    
        \begin{subfigure}[b]{0.24\textwidth} 
	    \begin{tikzpicture}[line width=0.8pt, =\tikzscale]
	\definecolor{gg}{RGB}{204,255,153}
	\definecolor{orang}{RGB}{255,192,182}
		\draw[draw = none, fill = gg] (0.0,0.0) rectangle (1.6,1.4);
	\draw[draw = none, fill = orang] (0.0,0.0) rectangle (-1.6,-1.4);
	\draw[->, line width = 1pt] (-1.7, 0) -- (1.7,0);
	\draw[->, line width = 1pt] (0, -1.6) -- (0,1.6);
	\draw[line width = 0.8pt] (-1.2, -1.2) -- (1.2,1.2);
	\draw (0.5,-0.4) node [anchor=west] {$U_{{\phi,i,t, V}}$};
	\draw (-0.1,1.4) node [anchor=east] {Voltage};
	\draw (-0.1,1.0) node [anchor=east] {imbalance};
	\draw (-0.1,0.6) node [anchor=east] {FSA component};
\end{tikzpicture}
        \caption[case_2]{Voltage imbalance component}    
        \label{fig:fsacomponent3}
    \end{subfigure}
    \begin{subfigure}[b]{0.24\textwidth} 
	    \begin{tikzpicture}[line width=0.8pt, =\tikzscale]
	\definecolor{gg}{RGB}{204,255,153}
	\definecolor{orang}{RGB}{255,192,182}
	\draw[draw = none, fill = gg] (0.0,0.0) rectangle (1.6,-1.4);
	\draw[draw = none, fill = orang] (0.0,0.0) rectangle (-1.6,1.4);
	\draw[->, line width = 1pt] (-1.7, 0) -- (1.7,0);
	\draw[->, line width = 1pt] (0, -1.6) -- (0,1.6);
	\draw[line width = 0.8pt] (-1.2, 1.2) -- (1.2,-1.2);
	\draw (0.5,0.2) node [anchor=west] {$U_{{\phi,i,t, I}}$};
	\draw (0,1.5) node [anchor=west] {Current};
	\draw (0,1.1) node [anchor=west] {imbalance};
	\draw (0,0.7) node [anchor=west] {FSA component};
\end{tikzpicture}
        \caption[case_2]{Current imbalance component}    
        \label{fig:fsacomponent4}
    \end{subfigure}
   
    \caption{Convention and components for active power (P) FAS. The ramp down active power needs are denoted by orange color, and vice versa. Ramp down flexibility reduces net load. The reactive FAS is similar to FAS design for P. In case of Q, the region shaded in orange denotes the need for capacitive flexibility and green denotes inductive Q. }
    \label{fig:componentfle}
\end{figure}
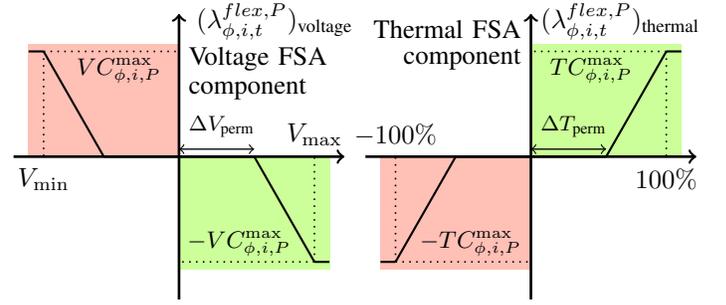
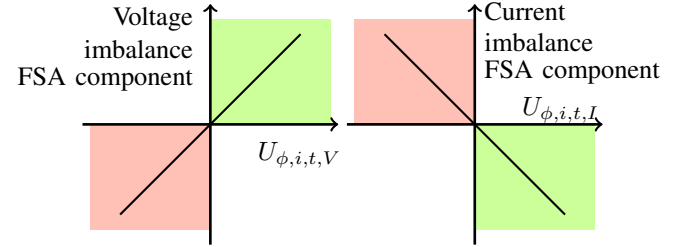

The voltage component of FAS for active and reactive power is governed by voltage magnitude.
The ramp down and ramp up voltage components of FAS for active power are given as
\begin{subequations}
\label{eq:flexibilityPrices}
\small{
\begin{equation} \begin{split}
    &  (\lambda_{\phi,i,t}^{\text{flex}_P+})_{\text{voltage}} =
 \mathbbm{1}_{(V_{\phi,i,t}\leq V_{\min})}\text{VC}^{\max}_{\phi,i,P}  + \\ & 
    \mathbbm{1}_{(V_{\phi,i,t}\in (V_{\min}, 1- \Delta V_{\text{perm}}))} \frac{\text{VC}^{\max}_{\phi,i,P} \Big(V_{\phi,i,t} - (1-\Delta V_{\text{perm}})\Big)}{\Big(V_{\min} - ( 1- \Delta V_{\text{perm}})\Big)} 
\end{split} \end{equation}
\vspace{-10pt}
\begin{equation} \begin{split}
    &(\lambda_{\phi,i,t}^{\text{flex}_P-})_{\text{voltage}} =   
    \mathbbm{1}_{(V_{\phi,i,t}\geq V_{\max})}(-\text{VC}^{\max}_{\phi,i,P})+ \\&
    \mathbbm{1}_{(V_{\phi,i,t}\in (1+ \Delta V_{\text{perm}}, V_{\max}))} \frac{(-\text{VC}^{\max}_{\phi,i,P}) \Big(V_{\phi,i,t} - (1+\Delta V_{\text{perm}})\Big)}{\Big(V_{\max} - ( 1+ \Delta V_{\text{perm}})\Big)}
\end{split} \end{equation} 
} \end{subequations}
{where $\Delta V_{\text{perm}}$ denotes the dead-band across 1 per unit (pu) desired voltage at a node for which activation of flexibility is not needed.}
    {$(\lambda_{\phi,i,t}^{\text{flex}_P+})_{\text{voltage}}\geq 0$ and $(\lambda_{\phi,i,t}^{\text{flex}_P-})_{\text{voltage}}\leq 0$ denote the ramp down and ramp up FAS for voltage component. For $(\lambda_{\phi,i,t}^{\text{flex}_P+})_{\text{voltage}}=(\lambda_{\phi,i,t}^{\text{flex}_P-})_{\text{voltage}}= 0$ implies that the voltage is within permissible limits in the DN for the phase $\phi$, node $i$ and time $t$.}
See Fig.~\ref{fig:componentfle} (A) for graphical representation of voltage component of FAS.
Due to space constraints, the reactive component of FAS is not detailed. 


\subsection{Thermal component of FSA}
The thermal loading of a branch is projected on network nodes to obtain the thermal component of FAS. 
The projected thermal loading of a node $i$, phase $\phi$ and time $t$ is denoted as $T_{\phi,i,t}$.
{The flow convention of a radial DN is positive if the power flows from the substation to the end of the feeder. A reverse flow of power towards the substation is assumed negative.}
The ramp down and ramp up thermal component of FAS for active power are given as
\begin{subequations}
\label{eq:costflex}
\small{
\begin{equation} \begin{split}
    &  (\lambda_{\phi,i,t}^{\text{flex}_P+})_{\text{thermal}} =
    \mathbbm{1}_{(T_{\phi,i,t} \geq 100)}\text{TC}^{\max}_{\phi,i,j,P} + \\ & 
    \mathbbm{1}_{(T_{\phi,i,t} \in (\Delta T_{\text{perm}}, 100))}
    \frac{\text{TC}^{\max}_{\phi,i,P} \Big(T_{\phi,i,t} - \Delta T_{\text{perm}}\Big)}{\Big(100- \Delta T_{\text{perm}}\Big)}
\end{split} \end{equation}
\begin{equation} \begin{split}
    &(\lambda_{\phi,i,t}^{\text{flex}_P-})_{\text{thermal}} =   
    \mathbbm{1}_{(T_{\phi,i,t} \geq 100)}(-\text{TC}^{\max}_{\phi,i,P})  \\&+
    \mathbbm{1}_{(T_{\phi,i,t} \in (\Delta T_{\text{perm}}, 100) )}
    \frac{(-\text{TC}^{\max}_{\phi,i,P}) \Big(T_{\phi,i,t} - \Delta T_{\text{perm}}\Big)}{\Big(100- \Delta T_{\text{perm}}\Big)}
\end{split} \end{equation} 
} \end{subequations}
where $\Delta T_{\text{perm}}$ denotes permissible thermal loading below which activation of flexibility is not needed,
{$(\lambda_{\phi,i,t}^{\text{flex}_P+})_{\text{thermal}}\geq 0$ and $(\lambda_{\phi,i,t}^{\text{flex}_P-})_{\text{thermal}}\leq 0$ denote the ramp down and ramp up FAS for thermal component.}
See Fig.~\ref{fig:componentfle} (B) for graphical representation of thermal component of FAS for active power.

\subsection{Imbalance component of FSA}
\label{subimbalance}
In this work, we assume all three-phase loads are balanced and DN imbalance is caused by single-phase loads and generation. 
The risk of over-compensation is observed in \cite{yao2020mitigating} for imbalance mitigation in DN. It should be considered that our RA does not increase the imbalance in the reverse direction by overcompensation. This could be ensured by distributed proportional control, as proposed in \cite{yao2020mitigating} or by formulating imbalance mitigation as an optimization problem.
Prior works, \cite{brandao2016centralized, girigoudar2020impact, nejabatkhah2017flexible, zeng2019system}, use a centralized optimization based approach for mitigating DN imbalances.
Authors in \cite{brandao2016centralized} propose a centralized master-slave controller for mitigating active and reactive power imbalance, where resources share proportional compensation roles. 
The centralized optimization is proposed in \cite{nejabatkhah2017flexible, zeng2019system} for avoiding the risk of overcompensation, which could happen with passive ways of imbalance compensation. \cite{nejabatkhah2017flexible} minimizes the negative and zero sequence currents and \cite{zeng2019system} utilize active power losses in DN as their objective function for the optimization.
We utilize the imbalance component in TPU-RA objective function while considering DN imbalances for FAS design.

Authors in \cite{girigoudar2020impact} compare unbalance indicators used in literature. They observe minimizing DN losses could also lead to a reduction in DN imbalances. 
The true imbalance definition requires the measurement of negative and positive sequence components of voltages and currents. Alternatively, \cite{bajo2015voltage} proposed the use of phase voltage and current magnitudes for measuring DN imbalances. These indices are evaluated in \cite{douglass2016voltage} and more recently in \cite{yao2020mitigating, yao2020overcoming}. 
{In our recent work, \cite{hashmi2022evaluation}, we have observed that magnitude based imbalance indicators are highly correlated with true imbalance definition based on sequence components for DNs.}
Next, we describe the magnitude-based indicators introduced in \cite{bajo2015voltage} for generating nodal signals for controlling single-phase flexible resources.

\subsubsection{Metric for voltage imbalance indicators}
Phase imbalance components for FAS are calculated based on instantaneous phase voltage and current magnitude imbalance.
Nodal voltage imbalance is defined using phase voltage unbalance rate (PVUR).
PVUR for node $i$ at time $t$ and phase $\phi$ is denoted as 
$
    (\text{PVUR})_{\phi,i,t} =V_{\phi,i,t}/ \bar{V}_{i,t},
$
where $\bar{V}_{i,t}=\frac{1}{3} \sum_{\phi \in \{A,B,C\}} V_{\phi,i,t}$ denotes the mean voltage at node $i$, time $t$.
Normalized voltage imbalance at node $i$ at time $t$ and phase $\phi$ is given as
\begin{equation}
    U_{{\phi,i,t, V}} = 1 -  (\text{PVUR})_{\phi,i,t},
\end{equation}

The imbalance for voltage and current ignores the phase angle of voltage and current magnitude measurements, similar to the model presented in \cite{douglass2016voltage}. 
\subsubsection{Metric for current imbalance indicators}
The current unbalance is also considered in FAS design, as voltage unbalance can create a current unbalance 6-10 times the magnitude of voltage unbalance \cite{x1pge}.
The current similar to thermal loading needs to be projected on nodes for calculating nodal signals.
Branch current between nodes $i$ and $j$ for phase $\phi$ at time $t$ is denoted as $I_{\phi,i,j,t}$.
The projected nodal current for phase $\phi$, node $i$ and time $t$ is given as
$
    I_{\phi,i,t} = \sum_j \zeta_{{i,j}} |I_{\phi,i,j,t}|,
$ {where $\zeta_{i,j}$ denotes the flow convention indicator calculated using \eqref{eq:flowdirections}.}
Nodal phase current imbalance rate (NPCUR) for node $i$ at time $t$ and phase $\phi$ is denoted as 
$
    (\text{NPCUR})_{\phi,i,t}= I_{\phi,i,t} / \bar{I}_{i,t},
$

and current imbalance at node $i$ at time $t$ and phase $\phi$ is
\begin{equation}
    U_{{\phi,i,t, I}} = 1 -  (\text{NPCUR})_{\phi,i,t}.
\end{equation}

\subsubsection{Imbalance component of FASs}

We present an example for identifying the direction of
the imbalance component of FAS
based on voltage imbalance. Consider voltages at phases A, B and C are 1.1, 1.05 and 1.03 respectively. The voltage imbalances are -0.04/1.06, 0.01/1.06 and 0.03/1.06 for phases A, B and C respectively.
In the case of phase A, ramp up flexibility needs to be activated which will increase the load supplied for phase A, in effect reducing the voltage at that phase. Similarly, ramp down flexibility needs to be activated in phases B and C to reduce load. This implies $U_{{\phi,i,t, V}} > 0$ will require activation of ramp down flexibility and vice versa.
The direction of activation for current imbalance can be decided in the same manner as for voltage imbalance described above. If $U_{{\phi,i,t, I}} < 0$ then ramp up flexibility activation is required 
and vice versa.
The imbalance component of FAS for P and Q flexibility for phase $\phi$, time $t$, node $i$ are defined as
\begin{subequations}
\begin{equation} 
  (\lambda_{\phi,i,t}^{\text{flex}_P+})_{\text{imb}} = \mathds{1}_{(U_{{\phi,i,t, V}} >0)} U_{{\phi,i,t, V}}  +  \mathds{1}_{(U_{{\phi,i,t, I}} <0)} U_{{\phi,i,t, I}} , 
\end{equation}
\begin{equation}
       (\lambda_{\phi,i,t}^{\text{flex}_P-})_{\text{imb}} = \mathds{1}_{(U_{{\phi,i,t, V}} <0)} U_{{\phi,i,t, V}}  +  \mathds{1}_{(U_{{\phi,i,t, I}} >0)} U_{{\phi,i,t, I}}, 
\end{equation}
\begin{equation}
    (\lambda_{\phi,i,t}^{\text{flex}_Q+})_{\text{imb}}  = \mathds{1}_{(U_{{\phi,i,t, V}} >0) }U_{{\phi,i,t, V}},
    \label{eq:currentimbalance1}
\end{equation}
\begin{equation}
    (\lambda_{\phi,i,t}^{\text{flex}_Q-})_{\text{imb}}  = \mathds{1}_{(U_{{\phi,i,t, V}} <0) }U_{{\phi,i,t, V}}.
    \label{eq:currentimbalance2}
\end{equation}
\end{subequations}
Note the direction of activation for reactive power flexibility for current imbalance mitigation is not obvious, therefore, \eqref{eq:currentimbalance1} and \eqref{eq:currentimbalance2} only considers the voltage imbalance component.

\subsubsection{Objective function components for imbalance mitigation}
For avoiding overcompensation of DN imbalances, we propose an imbalance component of the objective function for TPU-RA. This component is given as
\begin{equation}
        C_{V}^{\text{imbal}} = G_V \sum_t \sum_i \sum_{\phi} |V_{\phi,i,t} - \bar{V}_{i,t}|,
        \label{eq:voltageimbalance}
\end{equation}

where 
$G_V$ denotes the gain associated to 
the imbalance component of the objective function.
In \eqref{eq:voltageimbalance}, the absolute value function is not strictly convex. 
Convexification of \eqref{eq:voltageimbalance} is performed next.
Denote $\theta_{\phi,i,t} = |V_{\phi,i,t} - \bar{V}_{i,t}|$. This transformation will require inclusion of the following constraints \cite{optimizationArticle}:
\begin{subequations}
\label{eq:imbalance_obj}
\begin{equation}
    V_{\phi,i,t} - \bar{V}_{i,t} \leq \theta_{\phi,i,t},
\end{equation}
\begin{equation}
    - V_{\phi,i,t} + \bar{V}_{i,t} \leq \theta_{\phi,i,t}.
\end{equation}
\end{subequations}
\vspace{-20pt}
\begin{table}[!htbp]
	\scriptsize
	\caption {Voltage Unbalance Factor (VUF) limits} 
	\label{vuflimits}
	\vspace{-5pt}
	\begin{center}
		\begin{tabular}{ c | c}
			\hline
			Utility/Standard & VUF Limit\\
			\hline
			PG\&E \cite{x1pge} & 2.5\%\\
			NEMA MG-1-1988 \cite{nema}  & 1\%\\
			BC Hydro - Standard (\textit{Rural}) Unbalance \cite{r7bchydro} & 2\% (\textit{3\%})\\
			Europe EN 50160 - LV and MV (\textit{HV}) \cite{euunbalance} & 2\% (\textit{1\%})\\
			\hline
		\end{tabular}
		\hfill\
	\end{center}
\end{table}	
\vspace{-10pt}

\subsubsection{Tuning $G_V$ in multi-objective optimization setting}
\label{sectiontuning}
Note that the goal of resource activation is to keep imbalance within permissible limits and reduce it as much as possible. 
Table \ref{vuflimits} lists the voltage unbalance factor limits for different utilities, voltage levels and geographies. 
It is probable that fully compensating
network imbalance may require lots of flexibility to be activated.
However, there is no reason to fully compensate the network imbalance to zero.
Tuning of gain $G_V$ is discussed next.
In multi-objective optimization, all goals cannot be performed at optimal levels, as some goals may be in conflict with others. To prioritize different components of the multi-objective problem, Pareto optimality is used \cite{deb2005searching}.

\subsection{Flexibility activation signal}
FAS is a combination of voltage, thermal and imbalance components. The FAS for ramp down and ramp up active power flexibility are given as
\begin{subequations}
\begin{equation}
    \lambda_{\phi,i,t}^{\text{flex}_P+} = (\lambda_{\phi,i,t}^{\text{flex}_P+})_{\text{voltage}} + (\lambda_{\phi,i,t}^{\text{flex}_P+})_{\text{thermal}} + (\lambda_{\phi,i,t}^{\text{flex}_P+})_{\text{imb}} 
\end{equation}
\begin{equation}
    \lambda_{\phi,i,t}^{\text{flex}_P-} = (\lambda_{\phi,i,t}^{\text{flex}_P-})_{\text{voltage}} + (\lambda_{\phi,i,t}^{\text{flex}_P-})_{\text{thermal}} + (\lambda_{\phi,i,t}^{\text{flex}_P-})_{\text{imb}} 
\end{equation}
\end{subequations}
Due to space constraints equations for FAS for reactive power flexibility, i.e. $\lambda_{\phi,i,t}^{\text{flex}_Q+}$ and $\lambda_{\phi,i,t}^{\text{flex}_Q-}$ are not detailed.

\section{Optimization for resource activation}
\label{sectionOPT}
We detail the three-phase unbalanced resource activation (TPU-RA) in this section.
This formulation generalizes the balanced DN RA proposed in \cite{hashmi2021sest}.
The DSO activates distributed flexible and curtailable resources to avoid DN voltage, thermal and imbalance issues, which otherwise could have happened. The DSO aims to minimize the activation cost while reducing DN imbalance.
The bounds for flexible and curtailable resources are assumed to be known.
Next we detail the constraints and the optimization formulation.

\subsection{Flexibility definition}
The flexible resources are defined based on the ramp up and down, active (P) and reactive (Q) power levels. The ramp-up P flexibility increases nodal load and vice versa.
Q flex is defined as capacitive (+) and inductive (-).
{In this work, we assume that the flexibility ranges are known. The resource activation optimizations output the activated resources within these ranges given as}
\begin{IEEEeqnarray}{llll}
\IEEEyesnumber \label{eq:flexibility} \IEEEyessubnumber*
    \Delta P^{\text{flex}+}_{\phi,i,t} & \in & [0, P^{\text{flex}}_{\max,\phi,i,t}], & (\text{P injection or ramp down}),\\
    \Delta P^{\text{flex}-}_{\phi,i,t} & \in & [P^{\text{flex}}_{\min,\phi,i,t}, 0],  & (\text{P consumption or ramp up}),\\
    \Delta Q^{\text{flex}+}_{\phi,i,t} & \in & [0, Q^{\text{flex}}_{\max,\phi,i,t}],  & (\text{Q injection}),\\
     \Delta Q^{\text{flex}-}_{\phi,i,t} & \in & [Q^{\text{flex}}_{\min,\phi,i,t}, 0],  & (\text{Q consumption}).
\end{IEEEeqnarray}
Thus, $\Delta P^{\text{flex}}_{\phi,i,t} = \Delta P^{\text{flex}+}_{\phi,i,t} + \Delta P^{\text{flex}-}_{\phi,i,t}$ and $\Delta Q^{\text{flex}}_{\phi,i,t} = \Delta Q^{\text{flex}+}_{\phi,i,t} + \Delta Q^{\text{flex}-}_{\phi,i,t}$.
Each flexibility components for ramp up and ramp down P and Q have an associated cost value denoted as $\lambda_{\phi,i,t}^{\text{flex}_P+},\lambda_{\phi,i,t}^{\text{flex}_P-}, \lambda_{\phi,i,t}^{\text{flex}_Q+},\lambda_{\phi,i,t}^{\text{flex}_Q-}$ respectively.
For cases where $\lambda_{\phi,i,t}^{\text{flex}_P+},\lambda_{\phi,i,t}^{\text{flex}_P-}$ are both zero implying voltage and line loadings are within permissible bounds, both $\Delta P_{\phi,i,t}^{\text{flex}+}$ and $\Delta P_{\phi,i,t}^{\text{flex}-}$ should be zero. 
Similarly, for cases where $\lambda_{\phi,i,t}^{\text{flex}_Q+},\lambda_{\phi,i,t}^{\text{flex}_Q-}$ are both zero implying voltage within permissible bounds, $\Delta Q_{\phi,i,t}^{\text{flex}+}$ and $\Delta Q_{\phi,i,t}^{\text{flex}-}$ should be zero. 
In the absence of the above conditions being considered, the power balance constraint in TPU-RA implementation will not accurately represent the DN. This problem can be solved by introducing an integer variable in the TPU-RA or by redefining the flexibility constraint in \eqref{eq:flexibility} as
\begin{IEEEeqnarray}{lllll}
\IEEEyesnumber\label{eq:flexibility2} \IEEEyessubnumber*
    \Delta P^{\text{flex}+}_{\phi,i,t} &\in& [0, z_1 P^{\text{flex}}_{\max,\phi,i,t}] &=& [0,  P^{\text{flex}N}_{\max,\phi,i,t}],\\
    \Delta P^{\text{flex}-}_{\phi,i,t} &\in& [z_2 P^{\text{flex}}_{\min,\phi,i,t}, 0] &=& [ P^{\text{flex}N}_{\min,\phi,i,t}, 0], \\
    \Delta Q^{\text{flex}+}_{\phi,i,t} &\in& [0, z_3 Q^{\text{flex}}_{\max,\phi,i,t}] &=& [0,  Q^{\text{flex}N}_{\max,\phi,i,t}], \\
     \Delta Q^{\text{flex}-}_{\phi,i,t} &\in& [z_4 Q^{\text{flex}}_{\min,\phi,i,t}, 0] &=& [ Q^{\text{flex}N}_{\min,\phi,i,t}, 0],
\end{IEEEeqnarray}
where $z_1, z_2, z_3, z_4$ denotes binary variables. These binary variables are calculated as
\begin{equation} \begin{split}
    z_1 = \mathbbm{1}_{( \lambda^{\text{flex}P+}_{\phi,i,t} \ne  0)}, ~~~
    z_2 = \mathbbm{1}_{( \lambda^{\text{flex}P-}_{\phi,i,t} \ne  0)}, \\
    z_3 = \mathbbm{1}_{( \lambda^{\text{flex}Q+}_{\phi,i,t} \ne  0)}, ~~~
    z_4 = \mathbbm{1}_{( \lambda^{\text{flex}Q-}_{\phi,i,t} \ne  0)}.
    \label{eq:binaryvariables}
\end{split} \end{equation}
Since the activation signals are calculated prior to solving the resource dispatch optimization problem, 
thus flexibility limits calculated as in \eqref{eq:flexibility2} avoids the use of binary variables in the proposed TPU-RA, detailed next.


\subsection{Load and generation curtailment}
The load and generation curtailment cost is set at a level higher than the highest value of the FAS. This will ensure curtailment of load and generation are avoided if flexibility activation can solve network congestion and/or imbalance issues.
In order to ensure a feasible solution of TPU-RA, the limits for generation and load curtailment are defined as
\begin{equation}
  0 \leq \Delta P^{G}_{\phi,i,t} \leq P^{g}_{\phi,i,t}, ~~  \forall i \in \mathscr{N_G}, \forall \phi, 
  \label{eq:gencurt}
\end{equation}
\begin{equation}
  0 \leq \Delta P^{\text{curt}}_{\phi,i,t} \leq P^{d}_{\phi,i,t}, ~~  \forall i \in \mathscr{N_L}, \forall \phi.
  \label{eq:loadCurtailmentLimits}
\end{equation}

\subsection{Optimization formulation}

The objective of the resource dispatch problem is to minimize the cost of flexibility activation and load and/or generation curtailment over a time horizon under consideration, and also reduce the imbalance component described in \eqref{eq:voltageimbalance}.
For tractable implementation of TPU-RA, FASs 
are generated a priori to implementing the main optimization discussed earlier.
This decomposition of the optimization 
avoids the use of 
binary variables in the proposed formulation.
The decision variables for the optimization are $\Gamma = \{P_{\phi,j,t}^g,\Delta P^{\text{flex}}_{\phi,i,t}, \Delta Q^{\text{flex}}_{\phi,i,t},$ $ \Delta P^{\text{curt}}_{\phi,i,t}, \Delta P^{\text{G}}_{\phi,i,t} \}$ which denote active power flexible resource activated, reactive power flexible resource activated, active power generation curtailment and load shedding, respectively.
The objective function for time $t$ is given as
\begin{equation}\begin{split}
\label{eq:objectivefun}
   & \sigma(\lambda_{\phi,i,t}^{\text{flex}_P+},\lambda_{\phi,i,t}^{\text{flex}_P-}, \lambda_{\phi,i,t}^{\text{curt}_P}, \lambda_{\phi,i,t}^{\text{curt}_G} ) =  G_V \sum_i \sum_{\phi} \theta_{\phi,i,t}+
    \\& 
     \sum_{i \in \mathscr{N} } \sum_{\phi } \lambda_{\phi,i,t}^{\text{flex}_P+} \Delta P^{\text{flex}+}_{\phi,i,t} + 
     \sum_{i \in \mathscr{N} } \sum_{\phi } \lambda_{\phi,i,t}^{\text{flex}_P-} \Delta P^{\text{flex}-}_{\phi,i,t} \\ &  +
     \sum_{i \in \mathscr{N} } \sum_{\phi } \lambda_{\phi,i,t}^{\text{flex}_Q+} \Delta Q^{\text{flex}+}_{\phi,i,t} +
     \sum_{i \in \mathscr{N} } \sum_{\phi } \lambda_{\phi,i,t}^{\text{flex}_Q-} \Delta Q^{\text{flex}-}_{\phi,i,t} + \\ &
     \sum_{i \in \mathscr{N} } \sum_{\phi } \lambda_{\phi,i,t}^{\text{curt}_G} \Delta P^{\text{G}}_{\phi,i,t}+ 
     \sum_{i \in \mathscr{N} } \sum_{\phi } \lambda_{\phi,i,t}^{\text{curt}_P} \Delta P^{\text{curt}}_{\phi,i,t}. \vspace{-6pt}
\end{split}\end{equation}

We can select the objective function parameter values as follows:
\begin{equation}
   0 \leq \max(\lambda_{\phi,i,t}^{\text{flex}_P+}, |\lambda_{\phi,i,t}^{\text{flex}_P-}|)
   <  \lambda_{\phi,i,t}^{\text{curt}_G}, \lambda_{\phi,i,t}^{\text{curt}_P}.
    \label{eq:condition}
\end{equation}
\eqref{eq:condition} ensures that no load shedding is performed before availing other options. 
The full nonlinear optimization formulation is denoted as three-phase unbalanced resource activation or TPU-RA is given as
\begin{subequations}
\begin{equation}\begin{split}
\label{eq:porgobjective}
   \underset{\substack{\Gamma}}{\text{min}}~ \sum_t \sigma(P_{\phi,i,t}^g, \rho, \lambda_{\phi,i,t}^{\text{flex}}, \lambda_{\phi,i,t}^{\text{curt}_P}, \lambda_{\phi,i,t}^{\text{curt}_G} ) 
\end{split}\end{equation} 
\text{subject to,} \eqref{eq:flexibility2}, \eqref{eq:gencurt}, \eqref{eq:loadCurtailmentLimits} and
\begin{equation}
   ~V_{\min}^i \leq |V_{\phi,i,t}| \leq V_{\max}^i, ~ \forall i \in \mathscr{N} , t \in \{1,..,T\}, \forall \phi,
  \label{eq:voltage}
\end{equation}
\begin{equation}\begin{split}
 &  (P^g_{\phi,i,t} - \Delta P^G_{\phi,i,t}) - (P^d_{\phi,i,t} - \Delta P^{\text{curt}}_{\phi,i,t} - \Delta P^{\text{flex}}_{\phi,i,t}) + j(Q^d_{\phi,i,t}\\ &   - \Delta Q^{\text{flex}}_{\phi,i,t}) = \sum s_{ij}^t,  \forall i,j \in \mathscr{N}, \forall \phi \in \{A,B,C\},
 \label{eq:powerbalance}
\end{split}\end{equation}
\begin{equation}
  |S_{\phi,ij}^t| < s_{\phi,ij}^{\max}, ~~\forall~ i,j \in \mathscr{N},  \forall \phi \in \{A,B,C\}    \label{const:thermal}, 
\end{equation}
\begin{equation}
  P^g_{\phi,i,t} \in [P^g_{\min,\phi,i},P^g_{\max,\phi,i}] , ~~\forall~ i \in \mathscr{N_G}, \forall \phi     \label{const:genlimit}, 
\end{equation}
\begin{equation}
  S_{ij}^t  = \textbf{Y}_{\phi,ij}^*V_{\phi,i,t}V_{\phi,i,t}^* - \textbf{Y}_{\phi,ij}^*V_{\phi,i,t}V_{j,t}^*, ~ \forall (i,j) \in E \cup E^R, 
  \label{eq:ohmslaw}
\end{equation}
\begin{equation}
  \angle (V_{\phi,i,t} V_{\phi,j,t}^*) \in [\theta_{\phi,ij}^{\min}, \theta_{\phi,ij}^{\max}], ~~\forall~ i,j \in \mathscr{N}, \forall \phi 
  \label{const:phase}
\end{equation}
\end{subequations}
\eqref{eq:voltage}, \eqref{const:thermal} and \eqref{const:phase} denote the voltage constraint for nodes, thermal constraint and phase angle constraints for branches, respectively. \eqref{eq:powerbalance} denotes the nodal balance of active and reactive power in the network.
\eqref{const:genlimit} denotes the generator output power limits and
\eqref{eq:ohmslaw} denotes Ohm's law. Flexibility limits for active and reactive ramp up and ramp down are denoted in
\eqref{eq:flexibility2}.
\eqref{eq:gencurt} and \eqref{eq:loadCurtailmentLimits} place limits on generation and load curtailment.



\section{Numerical result}
\label{sectionNUM}
\begin{figure}[!htbp]
	\center
	\includegraphics[width=3in]{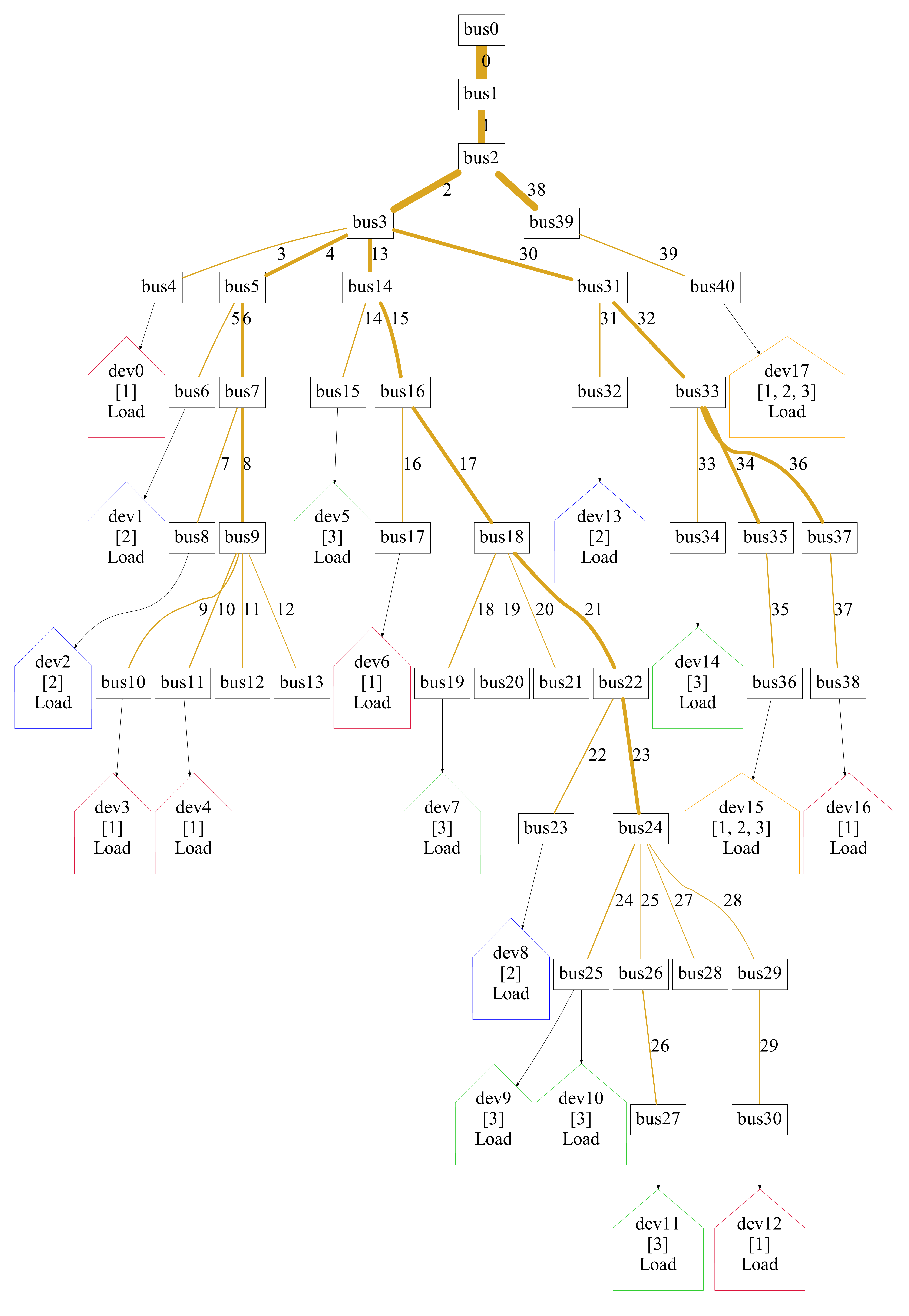}
	\vspace{-9 pt}
	\caption{\small{41 bus, 18 load (2 three-phase, 16 single phase load) DN}}
	\label{fig:network}
\end{figure}

The DN considered is an adaptation of one of the Spanish feeders described in \cite{koirala2020non}.
The reduced 3x3 primitive impedance matrix is used to include the effect of an isolated neutral conductor by the reduction proposed in~\cite{koirala2019impedance} to represent 4-wire European DN by its 3-wire equivalent.
The numerical experiment considers a 41 bus, 18 loads, low voltage distribution network.
Out of 18 loads, 16 are single-phase loads and two balanced three-phase loads.
All loads are connected to dedicated buses, except devices 9 and 10 are connected to bus 25.
The network is shown in Fig.~\ref{fig:network}.
The R/X ratio of the network is 6.85, typical LV network R/X network ranges from 2 to 10.

Table \ref{tab:consumerconnection} details the consumer phase connections and cumulative load (along with PV generation) details. All load nodes are having a 10 kWp single-phase PV generation also installed. Note for three-phase loads connected at nodes 36 and 40, the single-phase PV are connected in phase A. Cumulative installed PV in phase A is 80 kWp, in phase B is 40 kWp and phase C is 50 kWp.
The cumulative load as seen from the substation (node 0) are 312.4, 167.2, 145.6 kWh for phases A, B and C respectively.
The load and PV distribution imbalance are deliberately done to increase voltage and current imbalance in this test case. 
{The network details and load profiles used in this paper can be downloaded from GitHub repository, \cite{githubtpu}}.

\begin{table}[!htbp]
\caption{LV consumer connections and load in the test feeder}
\footnotesize
\vspace{-2pt}
\centering
\begin{tabular}{c|c|c|c|c|c|c|c}
\hline
     Consumer & Bus & \multicolumn{3}{c}{Phase connection} & \multicolumn{3}{c}{Cumulative energy [kWh]}    \\ 
     id& id & A& B& C& A& B& C\\ \hline \hline
     Dev 0	&4	&1	&0	&0	&-0.071	&0&	0\\
Dev 1	&6	&0	&1	&0	&0	&5.120&	0\\
Dev 2	&8	&0	&1	&0&	0&	9.684	&0\\
Dev 3	&10	&1	&0	&0	&131.88	&0	&0\\
Dev 4	&11	&1	&0	&0	&19.13	&0	&0\\
Dev 5	&15	&0	&0	&1	&0	&0	&50.90\\
Dev 6	&17	&1	&0	&0	&-18.28	&0	&0\\
Dev 7	&19	&0	&0	&1	&0	&0	&-0.198\\
Dev 8	&23	&0	&1	&0	&0	&-0.206	&0\\
Dev 9, 10	&25	&0	&0	&1	&0	&0&	-6.992\\
Dev 11	&27	&0	&0	&1	&0	&0	&-0.941\\
Dev 12	&30	&1	&0	&0	&184.7	&0	&0\\
Dev 13	&32	&0	&1	&0	&0	&48.59	&0\\
Dev 14	&34	&0	&0	&1	&0	&0	&19.593\\
Dev 15	&36	&1	&1	&1	&-30.47	&31.21	&24.97\\
Dev 16	&38	&1	&0	&0	&24.96	&0	&0\\
Dev 17	&40	&1	&1	&1	&0.725	&72.80	&58.24\\
\hline
\end{tabular}
    \label{tab:consumerconnection}
\end{table}

\begin{figure}[!htbp]
	\center
	\includegraphics[width=3.0in]{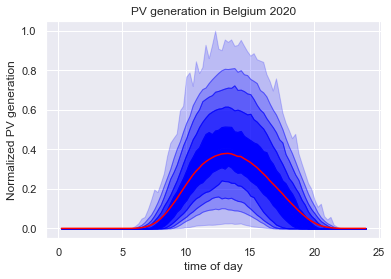}
	\vspace{-8pt}
	\caption{\small{PV generation profile in Belgium for 2020}}
	\label{fig:pvgen}
\end{figure}

The historical solar data is downloaded for Belgium from Elia's data portal \cite{EliaSolar}.
The normalized solar generation with respect to its capacity for the year is shown in Fig.~\ref{fig:pvgen}.
In this work, we consider normalized solar generation as the 70th percentile time series of the distribution plot shown in Fig.~\ref{fig:pvgen}.

\begin{figure}[!htbp]
	\center
	\includegraphics[width=3in]{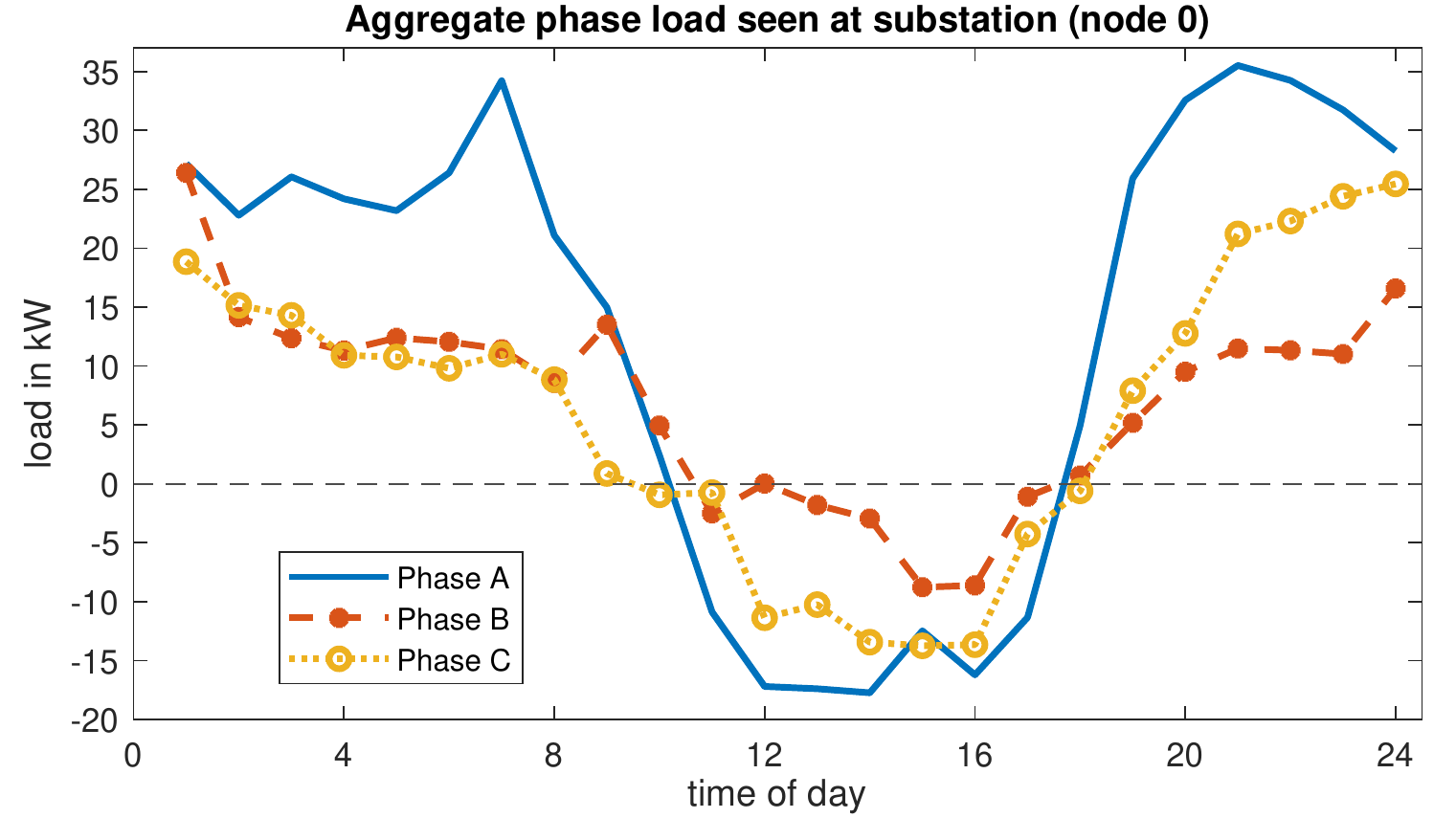}
	\vspace{-3pt}
	\caption{\small{Aggregate phase load}}
	\label{fig:phase}
\end{figure}

Fig. \ref{fig:phase} shows the aggregate phase load seen at the substation. Observe that the difference in active power load is high during the day when solar generation peaks and in the evening.

\begin{figure}[!htbp]
	\center
	\includegraphics[width=3.2in]{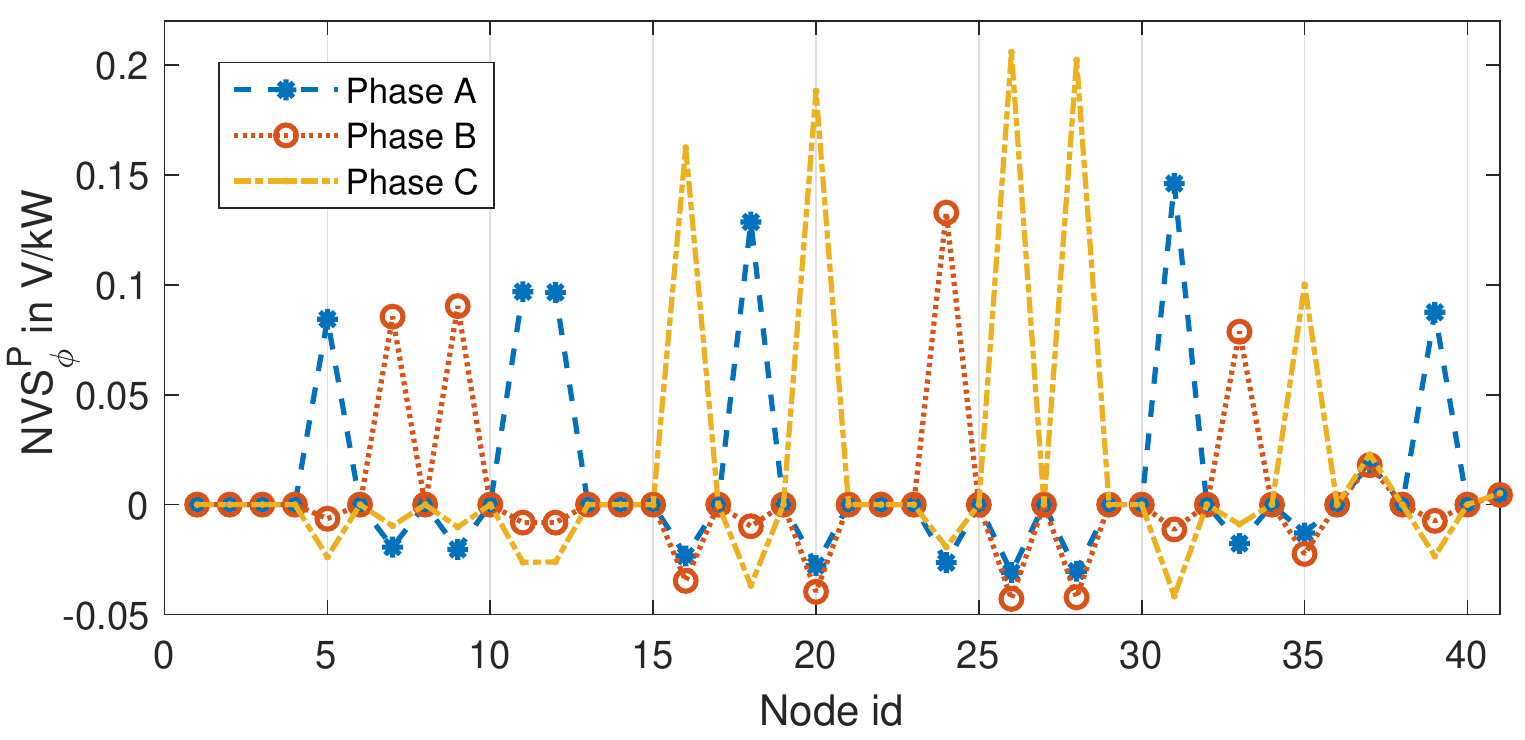}
	\vspace{-3pt}
	\caption{\small{Three-phase nodal voltage sensitivity}}
	\label{fig:sensitivity}
\end{figure}

The nodal voltage sensitivities are shown in Fig.~\ref{fig:sensitivity}.
As observed in \cite{weckx2016optimal}, the phase load change is expected to impact voltages in other phases. This is also observed in Fig.~\ref{fig:sensitivity}.

Using this network and load profiles, three-phase optimal power flows (OPF) are performed using PowerModelsDistribution.jl in Julia / JuMP
\cite{FOBES2020106664}.
Branch currents are calculated using a current-voltage formulation of AC-OPF \cite{ONeill2012}, and TPU-RA is executed using AC power flow model with polar bus voltage variables \cite{Cain2012}.

\subsection{OPF duals vs flexibility activation signals}
The locational marginal prices (LMP) for an unbalanced DN are observed to be different for a given time for different phases \cite{weckx2015locational}.
The dual variables associated with the power balance equation of the optimal power flow problem are often used as LMPs \cite{conejo2005locational}. 
These dual variables are active only when the active power balance constraint is not satisfied, as 
{Karush–Kuhn–Tucker (KKT)} conditions are active in such a case.
\begin{figure}[t]
	\center
	\includegraphics[width=3.4in]{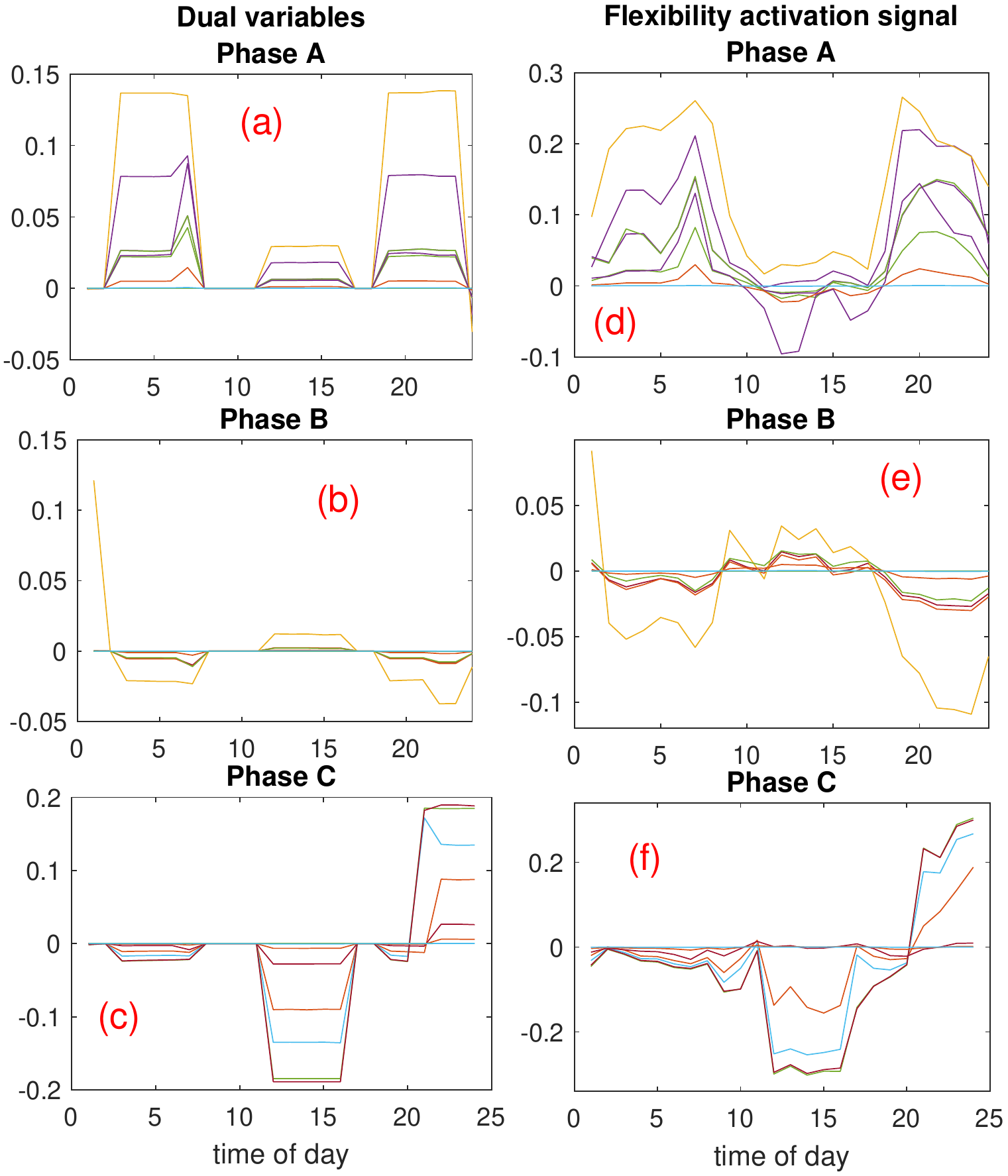}
	\vspace{-3pt}
	\caption{\small{OPF duals vs flexibility activation signals for phases.}}
	\label{fig:dualvsflex}
\end{figure}
The dual variables, however, do not provide corrective feedback before OPF constraint violation. Our proposed FAS hold some similarities with OPF dual variables as can be observed in Fig.~\ref{fig:dualvsflex}. Proposed flexibility activation signals unlike the OPF duals actively try to correct network flow and voltage levels if they exceed permissible safe levels of operation.


\subsection{Pareto optimal tuning of imbalance component gain}
\begin{figure}[t]
	\center
	\includegraphics[width=3.1in]{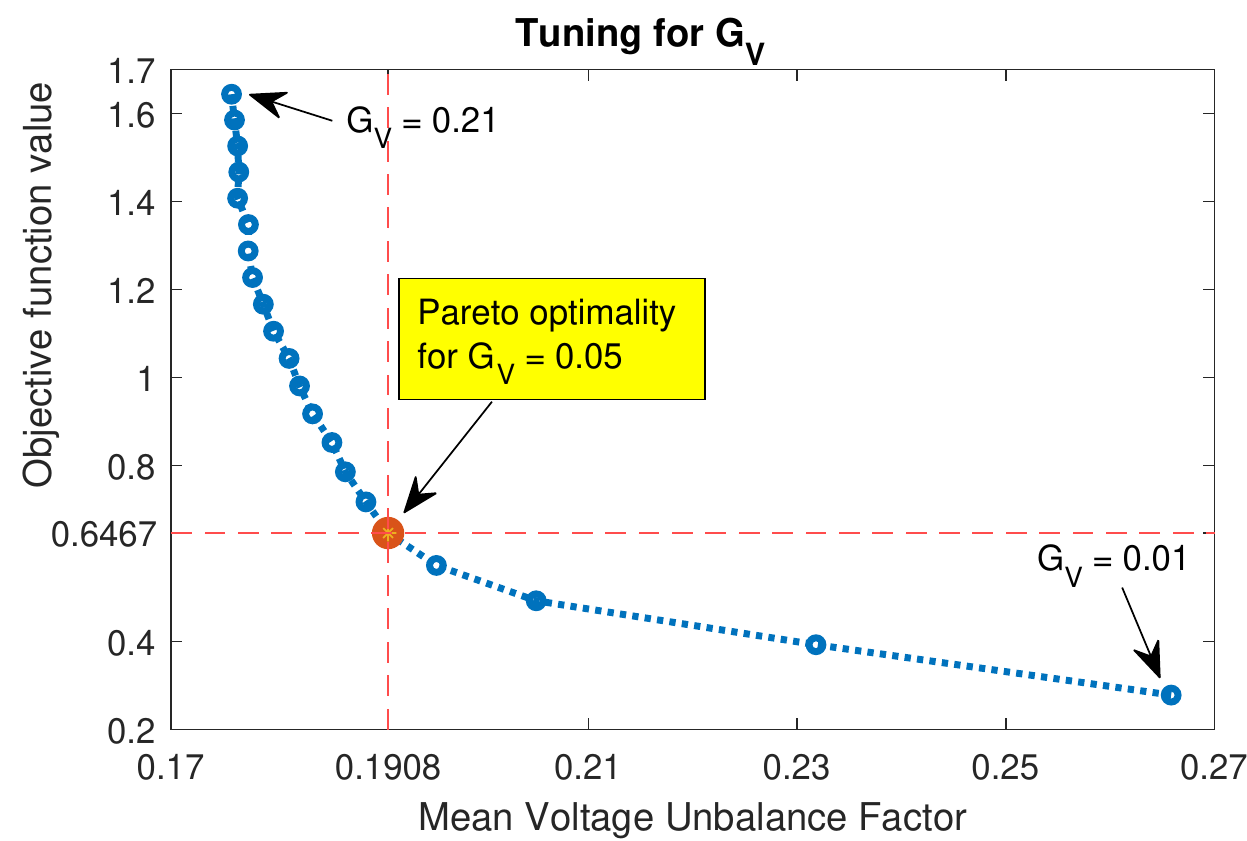}
	\vspace{-2pt}
	\caption{\small{Tuning $G_V$ using Pareto optimality.}}
	\label{fig:tuning}
\end{figure}

In section \ref{sectiontuning} we detailed the methodology to tune the imbalance component of the objective function of TPU-RA.
Fig.~\ref{fig:tuning} shows the Pareto optimal value of $G_V$ which reduces the voltage unbalance factor while considering objective function value.
For the tuned value of $G_V = 0.05$, we observe more than 81\% reduction in mean VUF and maximum VUF is reduced by more than 80\%.

\subsection{Simulation results }
\begin{figure}[!htbp]
	\center
	\includegraphics[width=2.9in]{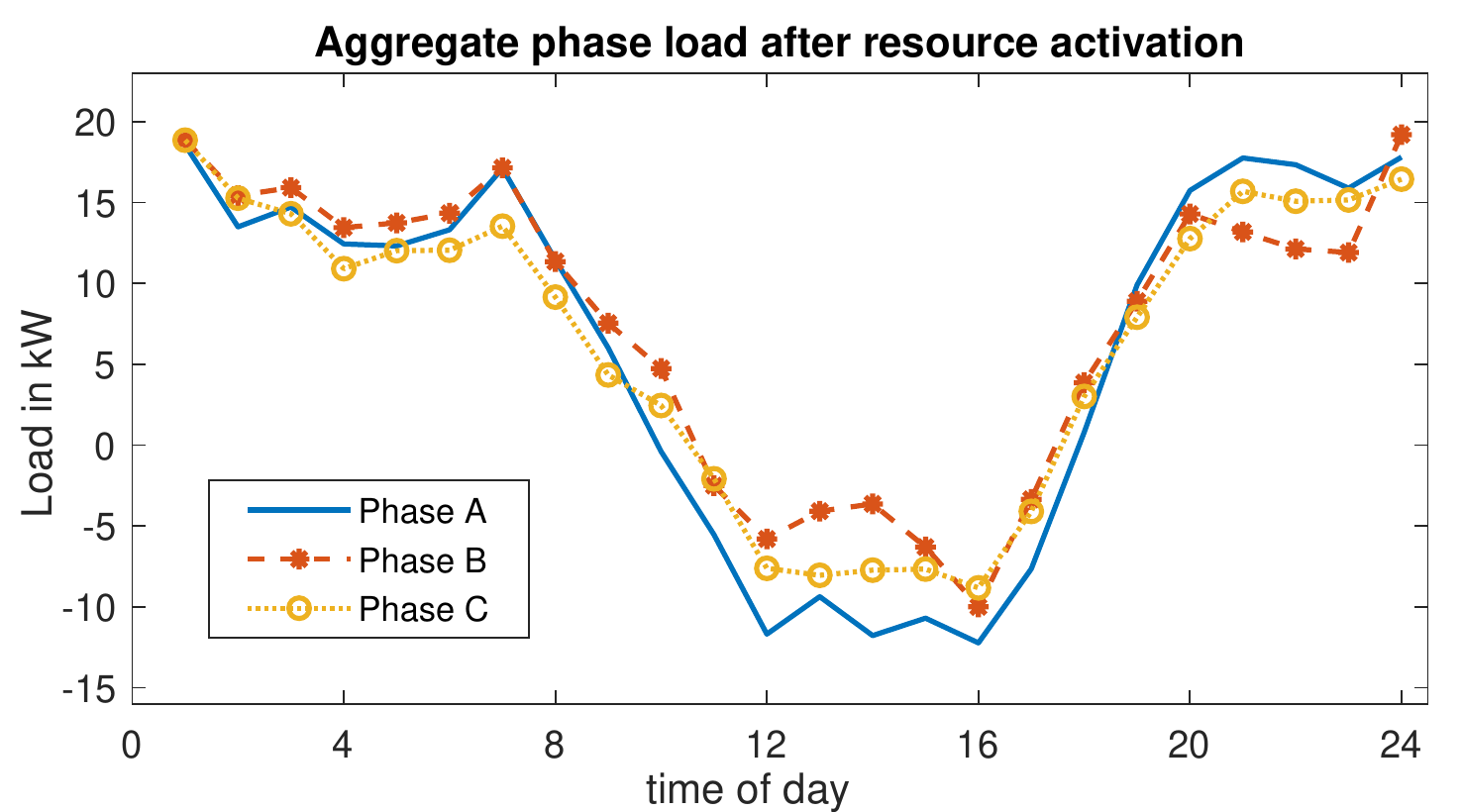}
	\vspace{-3pt}
	\caption{\small{Temporal phase load after resource activation.}}
	\label{fig:loadP}
\end{figure}

The voltage and thermal incidents observed for before and after resource activation are listed in Table \ref{tab:nominal}.
The goal of flexibility activation is to mitigate these network issues within permissible levels while minimizing the cost of operating such resources.
Highlighted items in Table \ref{tab:nominal} indicate hard bounds imposed by the DSO. In order to have a feasible OPF solution, the highlighted incidents should be zero.
Fig.~\ref{fig:loadP} shows the cumulative phase load after resource activation.

\begin{figure*}[!htbp]
	\center
	\includegraphics[width=7in]{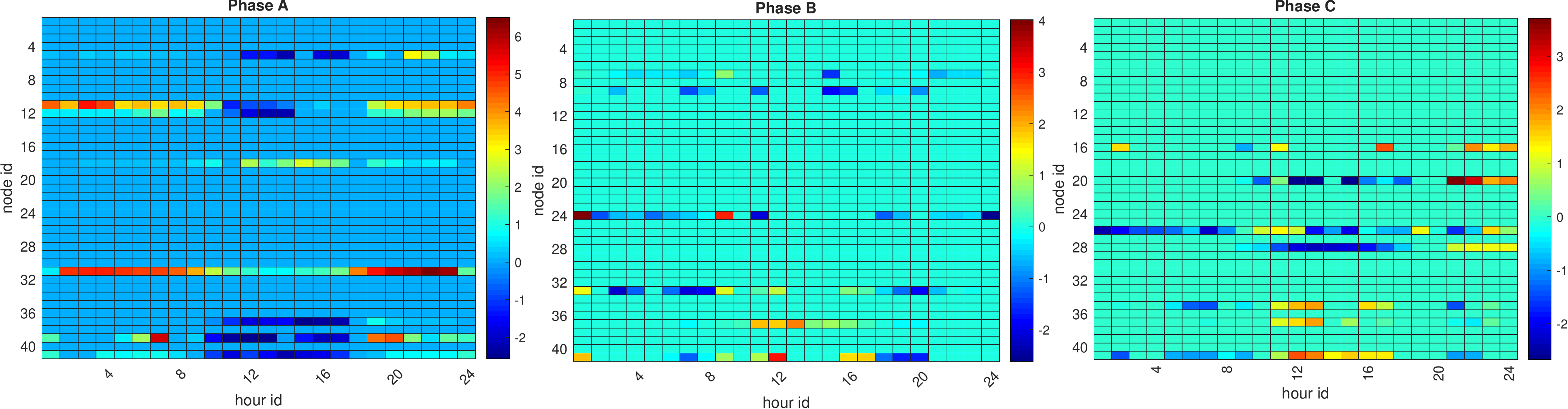}
	\vspace{-2pt}
	\caption{\small{Locational and temporal flexibility and curtailment power needs of the DN for phases A, B and C}}
	\label{fig:flexneed}
\end{figure*}

\begin{figure}[!htbp]
	\center
	\includegraphics[width=2.7in]{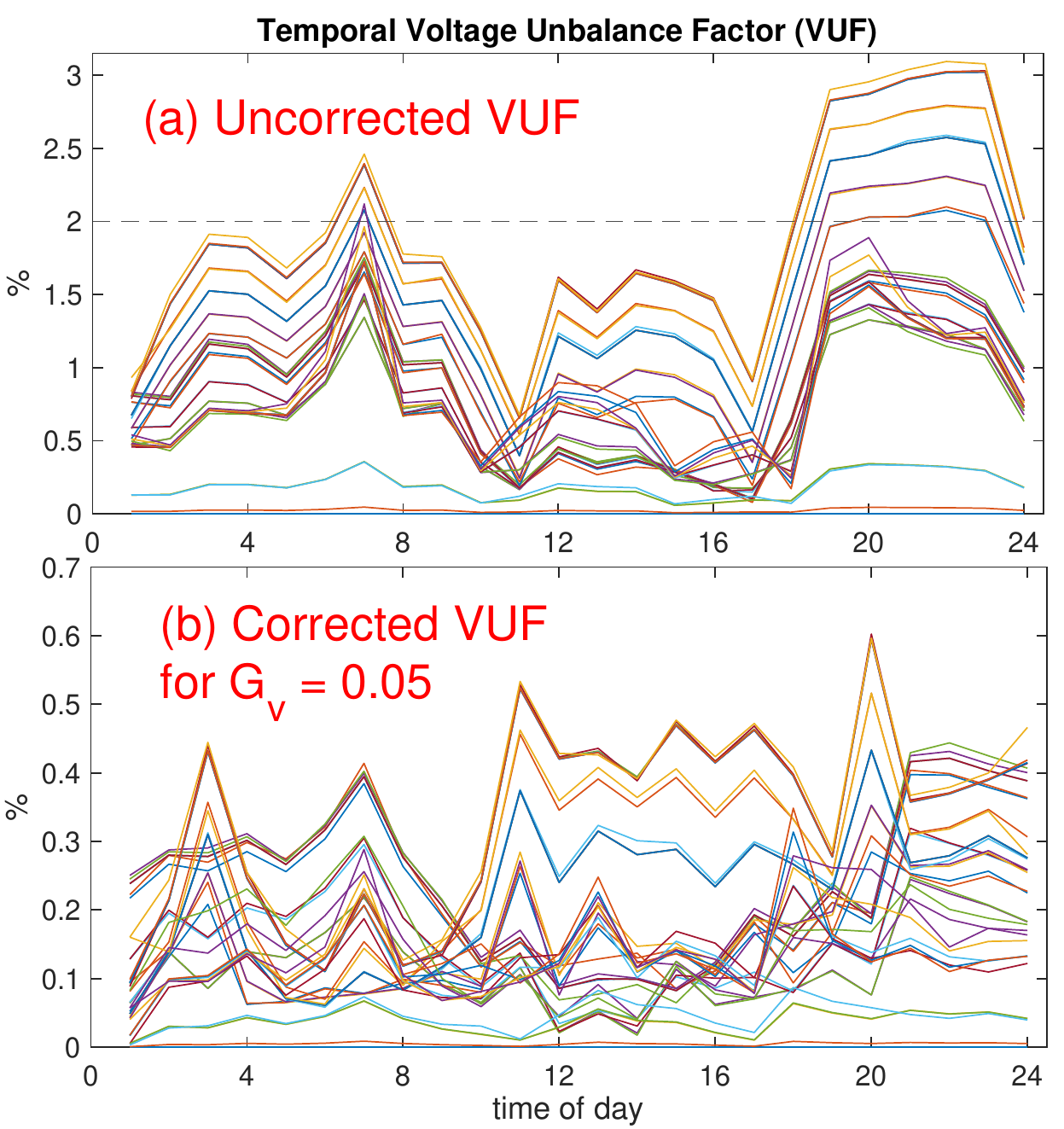}
	\vspace{-3pt}
	\caption{\small{Uncorrected and corrected temporal VUF comparison.}}
	\label{fig:vufplot}
\end{figure}

The temporal and locational flexibility ramp up and ramp down power needs of the DN considered is shown in Fig.~\ref{fig:flexneed}. {Fig.~\ref{fig:vufplot} shows the uncorrected and corrected temporal variation VUF of the DN. This VUF is calculated based on its true definition, i.e., the ratio of negative and positive sequence.}
The proposed framework can be utilized for feeder load reconfiguration to minimized resource activation even further.

\begin{table}[!htbp]
\caption{Uncorrected and corrected network incidents}
\vspace{-2pt}
\centering
\begin{tabular}{c|c|c|c|c}
\hline
     Network incident & \# instances & \# in \%   & \# instances & \# in \%   \\ \hline \hline
\hl{Under voltage} & 164 & 5.55\% & 0 & 0\%\\
Voltage below 0.96 pu & 530 & 17.95\%& 57 & 1.9\% \\
\hl{Over voltage} & 73 & 2.47\%& 0 & 0\% \\
Voltage above 1.03 pu & 280 & 9.48\%& 174 & 5.9\% \\
\hl{Thermal overload} & 24 & 0.8\%& 0 & 0\% \\
Mean VUF & 1.05 & -& 0.19 & -\\ 
Max VUF & 3.09 & -& 0.60 & - \\ \hline
\end{tabular}
    \label{tab:nominal}
\end{table}

\section{Conclusion}
\label{sectionCON}
We propose a resource activation (RA) framework in the context of an unbalanced distribution network.
This framework manages the operational states of curtailable and flexible resources for (a) mitigating nodal voltage issues, (b) mitigating branch over-loading, and (c) reducing voltage and current imbalances.
The resource activation is performed using a flexibility activation signal (FAS).
FAS design considers network states in absence of any RA.
The FAS resembles the OPF duals in the absence of RA. FAS, however, holds more network correction potential due to the droop-based design utilized for generating the FAS because of which it provides a non-zero value in case voltage and thermal loading exceeds their permissible levels prior to hard constraints of DN.
Based on the value of FAS, the limits of flexibility are pre-calculated before RA. This step avoids the use of binary variables in RA OPF.
The proposed three-phase unbalanced resource activation (TPU-RA) considers the cost of flexibility and curtailment activation, and voltage imbalance component. The gain associated with the voltage imbalance component is identified using Pareto optimality. Since voltage imbalance does not need to be minimized and only reduced within permissible limits, therefore, Pareto optimal gain will avoid over-activation of flexible and curtailable resources.
Numerical results show the efficacy of the proposed RA methodology.
For 41 bus typical Spanish DN, we observe the mean and maximum values of voltage unbalance factor are reduced by more than 80\% while ensuring all voltage and thermal network issues are resolved.
For a typical day, we quantify the temporal and locational needs for flexibility and curtailment.

In future work, we use the proposed framework for flexibility needs assessment for an unbalanced DN while considering future uncertainties. We observe that the framework can be adapted for feeder reconfiguration of connected phase for loads, for reducing DN voltage and current imbalance.
In this work, the reactive power flexibility is not considered, in future work we aim to quantify the impact of reactive power flexibility activation in the context of $3-\phi$ unbalanced DN.

\section*{Acknowledgement}
This work is supported by EUniversal project (\url{https://euniversal.eu/}) (Grant agreement ID: 864334) and the energy transition funds project BREGILAB organized by the FPS economy, S.M.E.s, Self-employed and Energy.
We also thank our EUniversal partners for their feedback and comments, and VITO for usage of their tool for generating Fig.~\ref{fig:network}.

\bibliographystyle{IEEEtran}
\bibliography{reference}

\appendix
\subsection{Analytical nodal sensitivity}
\label{appendix:nodalVs}

We use the \textit{perturb-and-observe} method, which approximates sensitivity components for the nodal voltage using small perturbations in active and reactive power and observe voltage magnitude changes. Nodal sensitivity calculation is detailed in Algorithm~\ref{alg:ns}.
Although nodal voltage sensitivities (NVS) calculations are computationally intensive and depend on network size \cite{munikoti2020analytical}, we implement Algorithm \ref{alg:ns} only once.
Calculated NVS are used for designing voltage and thermal components of the proposed FAS.
NVS acts as the merit order, nodes with a high cumulative impact on the network are valued higher. 
Flexible resources located at a more volatile node with high voltage fluctuations will receive greater value for RA according to the proposed FAS. For example, in a radial DN, the nodes at the end of a feeder will have higher NVS.
\vspace{-5pt}

\begin{algorithm}

	\small{\textbf{Inputs}: P \& Q perturbation sets, network layout, load profiles}
	\begin{algorithmic}[1]
	    \State Perform power flow for load profiles and calculate voltage matrix $V_{\phi,i}^{\text{ref}}$  for phase $\phi$, node $i$,
	    \State Perturb nodal P, Q load and perform power flow to identify
	    perturbed voltage matrix ($V_{\phi,i}^{\text{pert}_{\Delta P}}$) for phase $\phi$, node $i$, 
		\State Compute $\Delta V_{\phi,i}^P = V_{\phi,i}^{\text{pert}_{\Delta P}} - V_{\phi,i}^{\text{ref}}$, $\Delta V_{\phi,i}^Q = V_{\phi,i}^{\text{pert}_{\Delta Q}} - V_{\phi,i}^{\text{ref}}$. 
		\State Calc $\text{NVS}_{\phi,i}^P =  \Delta V_{\phi,i}^P / \Delta P$, $\text{NVS}_{\phi,i}^Q = \Delta V_{\phi,i}^Q / \Delta Q$,
		\State To reduce impact of magnitude of perturbation, $\text{NVS}_{\phi,i}^P, \text{NVS}_{\phi,i}^Q$ are calculated for different perturbation levels and averaged.
		\State $\text{NVS}_{\phi,i}^P, \text{NVS}_{\phi,i}^Q$ is a row vector of NVS for each phase and node that is used for FAS design as proposed in \cite{hashmi2021sest}.
	\end{algorithmic}
	\caption{\texttt{Nodal voltage sensitivity (NVS)}}
	\label{alg:ns}
\end{algorithm}

\subsection{Nodal projection of branch parameter}
\label{sec:nodalloadingprojection}
The proposed FAS is designed to generate a nodal activation signal for distributed flexibility based on nodal voltage magnitude, branch loading, and voltage and current imbalances. 
The challenge here is to project branch parameters onto nodes so as these projected values can be used for generating FAS components.
We utilize the projection mechanism proposed in our earlier work \cite{hashmi2021sest}.
The first step in this direction is to identify the flow direction based on the convention used.
In the case of radial DN, this convention is assumed to be the power flow from the substation to prosumers.
Consider $b_{\phi,y,z,t}$ denotes a branch connecting node $y$ to $z$ in phase $\phi$ for time $t$. The flow direction is identified using nodal voltage magnitudes as
\begin{equation}
    \zeta_{b_{\phi,y,z,t}} =  \mathbbm{1}(V_{\phi,y,t} \geq V_{\phi,z,t}) - \mathbbm{1}(V_{\phi,z,t} > V_{\phi,y,t}).
    \label{eq:flowdirections}
\end{equation}
If there is a reverse power flow, i.e., DGs generate more power than power consumed by connected loads, the value of $\zeta=-1$ for that branch.
Some nodes may be connected to more than one branch, for which the nodal projection is denoted as
\begin{equation}
    \frac{\sum_{\text{lines connected}}\text{(line loading)~(flow convention)~(line rating)}}{\sum_{\text{lines connected}}\text{(line rating)}}.
    \label{eq:projectlineload}
\end{equation}

Similarly, nodal current projection is denoted as
\begin{equation}
     \frac{\sum_{\text{lines connected}}\text{(line current)~(flow convention)~(line rating)}}{\sum_{\text{lines connected}}\text{(line rating)}}.
    \label{eq:projectlinecurrent}
\end{equation}

\end{document}